\documentclass[prb,twocolumn,showpacs,superscriptaddress,footinbib]{revtex4}

\pacs{71.10.Pm,73.40.Gk, 05.30.Fk,73.63.Nm}

\newcommand{\beq}{\begin{equation}}
\newcommand{\eeq}{\end{equation}}
\newcommand{\beqn}{\begin{eqnarray}}
\newcommand{\eeqn}{\end{eqnarray}}

\newcommand{\slp}{\raise.15ex\hbox{$/$}\kern-.57em\hbox{$ \partial $}}
\newcommand{\lnA}{\raise.15ex\hbox{$/$}\kern-.57em\hbox{$A$}}

\usepackage{graphicx}
\usepackage{epsfig}
\usepackage{bm}

\begin{document}
\title{Non-perturbative approach to backscattering off a
dynamical impurity in 1D Fermi systems}
\author{Daniel G.\ Barci}\thanks{Regular Associate of the International Centre for Theoretical Physics, ICTP, Trieste, Italy}

\affiliation{Departamento de F\'\i sica Te\'orica,
Universidade do Estado do Rio de Janeiro,Rua S\~ao Francisco Xavier 524, 20550-
013,  Rio de Janeiro, RJ, Brazil}

\author{L. Moriconi}
\author{M. Moriconi}\thanks{Regular Associate of the International Centre for Theoretical Physics, ICTP, Trieste, Italy}
\affiliation{Instituto de F\'\i sica, Universidade Federal do Rio de Janeiro,\\
C.P. 68528, Rio de Janeiro, RJ -- 21945-970, Brazil}

\author{Carlos M.\ Na\'on}
\author{Mariano J.\ Salvay}
\affiliation{Instituto de F\'\i sica La Plata, Departamento de F\'\i sica,
Facultad de Ciencias Exactas, Universidad Nacional de La Plata, CC 67, 1900 La Plata, Argentina}
\affiliation{Consejo Nacional de Investigaciones Cient\'\i ficas y
T\'ecnicas, Argentina}

\date{March 16, 2005}

\begin{abstract}
We investigate the problem of backscattering off a time-dependent
impurity in a one-dimensional electron gas. By combining the
Schwinger-Keldysh method with an adiabatic approximation in order
to deal with the corresponding out of equilibrium Dirac equation, we
compute the total energy density (TED) of the system. We show how the
free fermion TED is distorted by the backscattering amplitude and the
geometry of the impurity.
\end{abstract}

\maketitle

The physics of tunneling through static barriers is a topic of great importance in the subject of correlated quasi-one dimensional electron transport. A considerable number of well-established results derived over the last few years constitute by now the standard knowledge in the field \cite{fisher}. Relevant applications of the theoretical findings comprise the behavior of strongly anisotropic physical systems such as organic conductors \cite{organic conductors}, charge transfer salts \cite{salts}, quantum wires \cite{quantum wires}, carbon nanotubes \cite{CNT} and quantum Hall junctions \cite{kim}.

More recently, some attention has been paid to the problem of electron transport through time-dependent localized perturbations in one dimensional correlated matter. The interest has been to a large extent driven by the possibility of pumping charge and spin into a conductor (or semi-conductor) by means of an induced effective time-dependent potential \cite{pump}. Actually, it is not difficult to devise examples of relevant experimental setups which would provide the realistic arena for time-dependent one-body interactions. A laser beam applied to a carbon nanotube is likely to produce a periodic deformation of the lattice structure, playing the role of an effective time-dependent impurity; a similar effect should be expected in the localization of optical phonons in topological defects. It is worth noting, furthermore, that the cleanest system for experimental investigation would be given, probably, by a quantum Hall bar with a time-dependent gate producing a harmonic backscattering between edge states.

Despite the possible technological advances that the control of charge and spin currents could bring up, we are faced with the fundamental question of what we could learn about electron correlations using time-dependent potentials as out-of-equilibrium probes. To answer this question, at least partially, more theoretical work is needed to understand the detailed dynamics induced by this type of perturbation.

A potentially interesting observable that characterizes a tunneling process is the energy resolved current $j(\omega)$.
In Ref. \onlinecite{Komnik}, the relevance of this quantity in the upper region of the spectrum (i.e., above the Fermi energy, $\omega > E_F$) was emphasized, if one is interested to get information about correlations in the leads. To leading order in the tunneling amplitude, the energy resolved current $j(\omega)$ is related to a simpler observable, namely, the electron energy distribution function $n(\omega)$, also referred to as the total energy density (TED) in the literature \cite{Palmer} (for a precise mathematical definition see Eq. (\ref{d}) below). In general grounds, $n(\omega)$ gives information about the perturbation of the ground state in the leads due to tunneling processes. For a non-correlated material, for instance, $n(\omega)$ should vanish above the Fermi surface. Any population of the spectrum above the Fermi energy is originated from a combined effect of correlations and  multi-particle tunneling, due, on its turn, to out of equilibrium processes \cite{Komnik}. In Ref. \onlinecite{Gogolin}, $n(\omega)$ was evaluated in a model of correlated one-dimensional fermions with a time-dependent impurity coupled to the electron density through a forward-scattering coupling. In Ref. \onlinecite{Nos} this model was analyzed by means of functional bosonization \cite{Nos1}, focusing, in particular, on the transients produced by turning on the oscillatory impurity strength.

It is important to notice, however, that backscattering effects are expected to be relevant
as a rule, in all but rather exceptional experimental settings \cite{Poncharal}. Recently \cite{Gefen}, effects of backscattering in a time-dependent ultralocalized impurity were studied perturbatively, suggesting a striking enhancement of the total current for special values of the Luttinger parameters. Although very interesting by itself, it is necessary to take some care with perturbative calculations of tunneling processes, since multiple tunneling events may be missed in the series expansions, specially when dealing with finite barriers. The problem of backscattering by dynamical impurities is usually a very difficult one. Some models, like the spinless Luttinger model with a delta-like impurity, can be solved exactly for the specific value of the Luttinger parameter $K=1/2$ \cite{Chamon3}. However, for general strengths of the electron-electron interaction and finite-ranged impurities, there are no available closed analytical solutions (even in the free case). Therefore,
it is of fundamental importance to address non-perturbative strategies in such a context.

In this work we develop a non-perturbative calculation for a
somewhat simple model, chosen to illustrate our method. We are
interested in studying the effects that the backscattering off an
extended dynamical impurity of width $a$, oscillating with
frequency $\Omega$, will have on the spectrum of a one-dimensional
fermion gas. For this purpose we have implemented an adiabatic
approximation which allows one to compute the TED in a
straightforward way. The time scale of the barrier oscillations is
given by $1/\Omega$, while on the other hand the traversal
 time for tunneling is given by $a/|v|$ \cite{Buttiker}, where $v$ is the velocity of fermions inside the barrier.
 Therefore, if  the traversal time is much smaller than the oscillation time
($a/|v| \ll 1/\Omega$), the tunneling  can be considered as taking
place through an essentially static barrier. Thus, the adiabatic
approximation just consists in the calculation of the spectrum or,
in general, any correlation function, in the limit $\Omega
a/|v|\to 0$, neglecting subleading corrections, of order $(\Omega
a/|v|)^2$. Interestingly, the adiabatic regime was recognized as
the relevant one in the context of charge quantization in pumping
processes \cite{Chamon2}. It is correct to state that in the
adiabatic limit it is possible to evaluate in an exact way any
fermionic correlation function without relying neither on a
perturbative expansion in the backscattering coupling constant nor
on the finite range of the dynamical barrier. This means, in
principle, that it is possible to capture multiple tunneling
processes and bound states, which are absent in the case of
ultra-local potentials ($\delta$-like potentials) and small values
of the coupling constants. In appendix (\ref{appendix}) we have applied the method to the exactly 
solvable case of pure forward scattering to explicitly show how the adiabatic approximation works.  

As the computational starting point, let us consider the following Hamiltonian, which
describes the interaction of spinless fermions with an external effective time-dependent
potential $V(x,t)$, responsible for backscattering transitions between right and left movers:

\beq
H=H_0+H_{\rm imp} \ , \
\label{H}
\eeq
where
\beq
H_0=i\; \hbar \; v_F  \int dx\;\; \left(\psi_R^{\dagger} \partial_x\psi_R-\psi_L^{\dagger}\partial_x\psi_L\right)
\ , \
\label{H_0}
\eeq
and
\beq
H_{\rm imp}=g_{b}\int dx\;\;\left(\psi^{\dagger}_R\psi_L+\psi^{\dagger}_L\psi_R\right)\;V(x,t)
\ . \
\label{H_b}
\eeq
Above, $g_{b}$ is the coupling constant associated to the backward scattering of electrons
caused by the presence of a time-dependent harmonic barrier. It is also possible to consider
a more general Hamiltonian with forward scattering couplings in addition to the backward
coupling considered here, however, this does not lead to any additional interesting physical
effect. Our results can be in fact extended without much difficulty to the case where a
forward scattering coupling is taken into account. 

Although we have verified that both the method and the general
results are independent of the explicit details of $V(x,t)$, we
will use, just to fix ideas, a square potential profile,

\beq
V(x,t)=(\Theta (x + a/2) - \Theta(x - a/2))\cos(\Omega t) \ , \
\label{V}
\eeq
where $a$ is the width of the square potential and $\Omega$ the oscillation frequency.

We are particularly interested in obtaining the TED for the above model. We recall that in
the Wigner representation the TED can be written in terms of the fermion correlation function
as
\begin{equation}
n(\omega,R,T) = -i \int^{\infty}_{-\infty} d\tau\;
e^{i \omega \tau}G_{+ -}(r = 0, R, T , \tau)  \ , \
\label{d}
\end{equation}
where we have introduced the closed time path formalism \cite{CTP}
in which fermion propagators are time-ordered along the usual
Schwinger-Keldysh time contour $C$:
\begin{eqnarray}\label{e}
G_C =  \left(\begin{array}{cl} G_{++} & G_{+-}
\\ \\ G_{-+} & G_{--} \end{array} \right) \ . \
\end{eqnarray}
The subscripts $+$ and $-$ refer to fields defined in the upper
and lower branches of $C$, respectively, corresponding to the forward
$(+)$ and backward $(-)$ time evolution. Above, $r, \tau$ and $R, T$ are the spatial and
temporal relative and center of mass
coordinates, respectively.


The static case, $\Omega=0$, is exactly solvable. 
It corresponds to take the limit in Eq. (\ref{d}) such that, 

\beq
n_{\rm static}(\omega)=\lim_{
\Omega\to 0 }n(\omega,R=0,T)
\label{static}
\eeq
where we choose to calculate $n_{\rm static}(\omega)$ at the center of the barrier $R=0$.

Physically, the above limit means that the TED is assumed to be probed in very short time scales, involving an external process 
with a relaxation time $T_0\ll 1/ \Omega$. In this regime, the TED behaves as effectively static and is given by Eq. (\ref{static}). 

On the other hand, if $T_0\gg 1/ \Omega$, the TED given by Eq. (\ref{d}) is a very rapid oscillating function of time and it is not the relevant quantity to evaluate. An arbitary observable in this regime should be computed by means of an averaging procedure. The meaningful physical observable in our case is, therefore, the averaged TED 

\begin{equation}
\overline{n}(\omega) = \frac{\Omega}{2\pi}\int^{2
\pi/\Omega}_{0}n(\omega,R=0,T) dT  \ , \
\label{h}
\end{equation}
It is important to notice that this averaged quantity is purely dynamical and is in general disconnected from the ``instantaneous'' 
or static regime. Although we can take the limit $\Omega\to 0$ in Eq. (\ref{h}), this does not correspond to the static limit since,  
the definition of the averaged TED implies that it should be probed over an infinite time interval. Mathematically, it is simple to 
understand from  Eqs.\ (\ref{static}) and (\ref{h}) that 

\beq
\lim_{\Omega\to 0}\overline{n}(\omega)\neq  n_{\rm static}(\omega) \ , \
\eeq
once we cannot interchange the limit with the integral operator.

Since we are interested in the dynamical aspects of the problem we concentrate ourselves in the calculation of 
the averaged TED given by Eq. (\ref{h}). For this purpose, we have calculated the Green's function and TED, Eq. (\ref{d}),
in the adiabatic approximation, which, for the potential given by
(\ref{V}) amounts to substituting $g_{b}\cos( \Omega t)\rightarrow
g_{b}$. Then, we have restored the temporal dependence and
evaluated the averaged TED using Eq. (\ref{h}).
The validity of this procedure ({\em i.\ e.\ }of restoring the time dependence in $n(\omega)$ after integrating the Green's function), 
relies in the extra condition   $\omega
\gg \Omega$, which allows a practical computation of the TED, and
preserves non-trivial physics, holding beyond perturbation theory. The relaxation of  this condition forces us to restore the time dependence at the Green's function level, turning  the calculation considerable more difficult.  
 Therefore, our results are useful in the frequency range $\omega \gg \Omega \gg 1/T_0$. The first inequality is imposed to simplify calculations, while the second one is implicit in the definition of the averaged TED, Eq. (\ref{h}).

\begin{figure}[tbph]
\hspace{-1.0cm}
~~~~~\includegraphics[width=8.4cm, height=7.3cm]{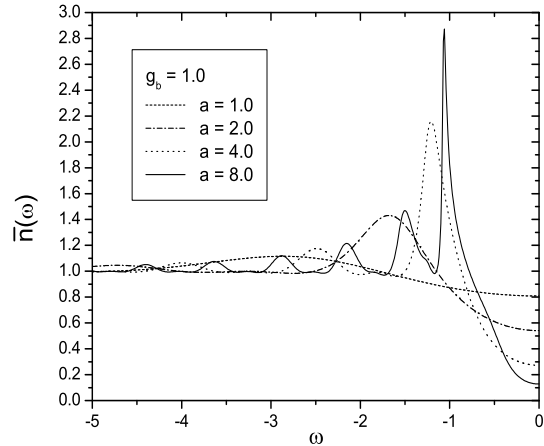}
\caption{Averaged energy density $\overline{n}(\omega)$ as a function of
the frequency $\omega$ for fixed backscattering coupling $g_{b}$ and variable potential range $a$.
$\omega$ and $g_b$ are measured in units of the Fermi energy $E_F=v_F k_F$ while the potential range $a$ is measured in units of the inverse of the Fermi momentum $1/k_F$.}
\label{fig1}
\end{figure}
\begin{figure}[tbph]
\hspace{-1.0cm} ~~~~~\includegraphics[width=8.4cm,
height=7.3cm]{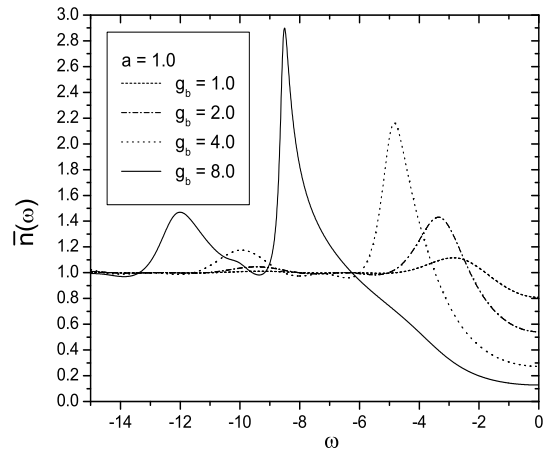} \caption{Averaged energy density
$\overline{n}(\omega)$ as a function of the frequency $\omega$ for
fixed potential range $a$ and variable backscattering coupling
$g_b$.} \label{fig2}
\end{figure}

Our main results are depicted in Figs. \ref{fig1} and \ref{fig2},
where $\overline{n}(\omega)$ is shown for fixed $g_{b}$, with
variable $a$ (Fig.\ref{fig1}), and for fixed $a$, with variable
$g_{b}$ (Fig.\ref{fig2}). Observe that at this level of approximation 
the curves are independent
of $\Omega$. It is worth noting, however, that the range of validity of
our predictions does depend on the external frequency due to the
condition $\omega \gg \Omega$. Also, in both cases, in contrast to
the situation in which only forward scattering is taken into
account \cite{Gogolin,Nos}, no gain or loss of energy-quanta of $n
\Omega$ takes place (see appendix \ref{appendix}). This is in fact a consequence of the $\omega
\gg \Omega$ regime considered here. However, one may clearly
observe that backscattering breaks the uniformity of the Fermi
sea, characteristic of the free electron gas. The TED
$\overline{n}(\omega)$ has its maximum peak for $\omega <
-|g_{b}|$. For $\omega > -|g_{b}|$, the TED shows a pronounced
decay, if compared to the free case behavior. When $|a g_{b}|$
grows, the peak tends to be situated at $\omega \approx -|g_{b}|$,
and the valley observed for $\omega > -|g_{b}|$ is drastically
lowered, indicating an important depletion in the population near
the Fermi surface (in this calculation we have tuned the Fermi
energy to $\omega=0$). Concerning the main peak, it is interesting
to note that the maximum value of the averaged TED obeys the power
law $max [\overline{n}(\omega)]\approx \sqrt{|a g_{b}|}$, for
$|ag_{b}|$ sufficiently large. Although the present model does not
contain correlations, it is clear from these results that the
presence of backscattering off the dynamical impurity will modify
correlations in a relevant way, mainly for large values of $|a
g_b|$, when the structure of the electronic density near the Fermi
surface is strongly affected. Let us emphasize again that the peak
structure we found (for the region $\omega \gg \Omega \gg 1/T_0$)
is not of static origin. Indeed, the static TED can be easily
computed (Eq. (\ref{static})) and it presents an oscillating 
structure with maxima of equal height.

We now sketch some technical details concerning the adiabatic
approximation. The main idea of this approximation, as stated
above, is that in a regime of sufficiently slow barrier
oscillations, $\Omega a/|v| \ll 1$, the spectrum of the system can
be considered essentially static. In this case, we can safely
consider (formally) $g_{b}\cos(\Omega t) \simeq g_{b}$.  The Dirac
equation corresponding to the Hamiltonian (\ref{H}) can be written
down as 
\beq 
i\partial_t \Psi = \left(H_0+H_{imp}\right) \Psi .
\label{dirac}
\eeq 

We seek, then, for stationary solutions of energy $E$.
Solving the resulting pair of coupled differential equations for
the right and left components of $\Psi$, we get for the right
field (to save notation we drop the index $R$):
\begin{equation}
\Psi_{E}(x) = \frac{1}{\sqrt{2\pi}}\frac{e^{-i a (E + k)/2}}{\frac{g_{b}}{E + k} + e^{-i a
 k}}\left(e^{i k x} + \frac{g_{b}\;e^{- i k x}}{E + k}\right) \ , \ \label{c}
\end{equation}
in the region $-a/2 < x < a/2$, where $ k^{2} = E^{2} - g_{b}^{2}$
(from now on, we take $\hbar=1, v_F=1$). We obtain a similar
result for the left moving fermion. Of course, outside the
barrier, where $V(x,t) = 0$, one has just plane wave solutions for
both chiral components. Inside the barrier, due to the presence of
backscattering, the solutions are plane waves or exponential
decays, depending on the sign of $ k^{2}$. This, in turn,
determines the energy regions in which one or the other kind of
wave function is defined.

In terms of the wave function obtained above, the right moving component of the Green's function reads:
\begin{eqnarray}
\lefteqn{G_{+ -}(R , r, T , \tau) = i
 \int^{0}_{-\infty}dE\;\times} \nonumber \\
&&\Psi^{*}_{E}(R - r/2, T - \tau/2)\Psi_{E}(R + r/2, T +\tau/2) \ . \ \label{f}
\end{eqnarray}
Substituting (\ref{c}) into (\ref{f}), using Eq. (\ref{d}) and explicitly evaluating the integrals, we obtain for
the TED at the center of the barrier the following expression:
\begin{widetext}
\begin{equation}
\label{g} n(\omega) =  \frac{( \Theta(
-\omega) - \Theta( -\omega - |g_{b}|))( g_{b} +
\omega)}{\cosh(\sqrt{g_{b}^{2} - \omega^{2}}a)\;\;g_{b} + \omega}
+ \frac{\Theta( -\omega - |g_{b}|)\;\;(g_{b} + \omega +
\sqrt{\omega^{2} - g_{b}^{2}})^{2}}{(\omega +  \sqrt{\omega^{2} -
g_{b}^{2}})^{2} + g_{b}^{2} + 2 g_{b} (\omega +  \sqrt{\omega^{2}
- g_{b}^{2}})\cosh(\sqrt{g_{b}^{2} - \omega^{2}}a)} +O(\Omega a)^2 \ . \ \label{pre-ted}
\end{equation}
\end{widetext}
The first term in the right hand side of (\ref{pre-ted}) comes
from the contribution of non-plane wave solutions ($|E|<|g_b|$),
while the second term comes from the energy region ($|E|> |g_b|$)
where the spectrum is composed essentially by plane waves. It is
interesting to note that no perturbation theory in the
backscattering amplitude $g_b$ is able to capture the physics of
the first term, since perturbation theory only deals with small
perturbation on plane waves.

Finally, restoring the temporal dependence through the formal substitution
$g_{b} \rightarrow g_{b}\cos( \Omega T)$, and performing the average over a period
(Eq. (\ref{h})), we obtain the plots shown in Figs. 1 and 2, discussed above.

To summarize, we have studied the effects of backscattering in a one dimensional fermion gas
with an oscillatory square barrier.
We have calculated the TED in the adiabatic approximation. This method allowed us to consider
finite barriers in a non-perturbative regime. In this way we were able to obtain the
dependence of TED with the coupling constant $g_b$ and with the geometry of the barrier, given essentially by its width $a$. We have found that the structure of the Fermi sea may be strongly modified by the presence of the dynamical barrier, producing a peak structure in the TED at energies around the typical backscattering energy $g_b$. Also, the probability of finding electrons near the Fermi surfaces may be drastically suppressed when the parameter $a g_{b}$ becomes large enough. We also found an interesting power law dependence of the maximum of the TED peak, given by  $max [\overline{n}(\omega)]\approx \sqrt{|a g_{b}|}$. This structure opens the interesting possibility of the experimental determination of microscopic quantities like the backscattering strength $g_b$ and  the effective width of the potential $a$. Of course, the spectrum modifications are expected to affect the electron correlations in the wire not trivially. We are currently analyzing the combined effect of electron-electron interactions in the wire with backscattering due to strong and extended dynamical barriers.

\acknowledgments

This work was partially supported by Universidad Nacional de La
Plata (Argentina), Consejo Nacional de Investigaciones Cient\' ificas
y T\'ecnicas, CONICET (Argentina), Universidade do Estado do Rio de
Janeiro (Brazil), Universidade Federal do Rio de Janeiro (Brazil).
The authors acknowledge support from CAPES (Brazil) and Fundaci\'on
Antorchas (Argentina) for a grant for scientific cooperation.
The Brazilian agencies CNPq and FAPERJ are also acknowledged for
financial support.

\appendix

\section{Forward Scattering}
\label{appendix}

In this section we consider an exactly solvable problem, in order to ilustrate
how the adiabatic approximation works. 

Let us consider the Hamiltonian 

\beq
H=H_0+H_{\rm imp} \ , \
\label{AH}
\eeq
where
\beq
H_0=i\; \hbar \; v_F  \int dx\;\; \left(\psi_R^{\dagger} \partial_x\psi_R-\psi_L^{\dagger}\partial_x\psi_L\right)
\ , \
\label{HA_0}
\eeq
and
\beq
H_{\rm imp}=g_{f}\int dx\;\;\left(\psi^{\dagger}_R\psi_R+\psi^{\dagger}_L\psi_L\right)\;V(x,t)
\ . \
\label{H_f}
\eeq
Above, $g_{f}$ is the coupling constant associated to the forward scattering of electrons caused by the presence of a time-dependent harmonic barrier.

In this case, the Dirac equation  (\ref{dirac}) is easily solved, since the right and left components of the spinor decouple. 
Then, the right mover component is given by a expression of the type $\psi_R\sim e^{i kx}$ while the left one reads $\psi_L\sim e^{-i kx}$.
Now, it is a simple matter to calculate the Green's function (Eq. 
(\ref{f})) and the averaged TED ( Eq.\ (\ref{h})) is given in the adiabatic approximation by,  

\begin{eqnarray}
\bar{n}(\omega)&=&\sum_{n=0}^{\infty}\sum_{i=0}^{2n}(-1)^i \frac{(2n)!}{(n!)^2(2n-i)! i!} \nonumber \\
&\times& \left( \frac{g_f a}{4} \right)^{2n} \Theta(-\omega-\Omega(n-i))
\label{fadiabatic}
\end{eqnarray}
where we have chosen $\hbar=1$ and $v_F=1$.

However, in this case, it is not necessary to invoke any type of approximation, since this problem is exactly solvable\cite{Gogolin,Nos,Nos1}. 
The reason behind this, is that this decoupling property between right and left movers is present in the full quantum problem and not merely in the static Dirac equation. The exact result for TED in the forward scattering case is given by

\begin{eqnarray}
\bar{n}(\omega)&=&\sum_{n=0}^{\infty}\sum_{i=0}^{2n}(-1)^i \frac{(2n)!}{(n!)^2(2n-i)! i!} \nonumber \\
&\times& \left( \frac{g_f \sin[\Omega a/4]}{\Omega} \right)^{2n} \Theta(-\omega-\Omega(n-i)) \ . \
\label{fexact}
\end{eqnarray}

Comparing Eqs.\ (\ref{fadiabatic}) and (\ref{fexact}), we clearly verify that the adiabatic approximation gives the correct result 
at leading order in  $\frac{\Omega a}{v_F}\ll 1$ (here,  we have recovered the original units).

\end{document}